\documentclass[trackchanges]{aastex62}

\received{}
\revised{}
\accepted{}
\submitjournal{ApJ}

\shorttitle{}
\shortauthors{}
\begin{document}

\title{Constraints on  gamma-ray and neutrino emission from  \textsc{NGC 1068} with the MAGIC telescopes}

\correspondingauthor{Alessandra Lamastra, Saverio Lombardi, Antonio Stamerra}
\email{alessandra.lamastra@inaf.it, saverio.lombardi@inaf.it, antonio.stamerra@inaf.it}

%\author[0000-0002-0786-7307]{author1....author2...}
%\affil{.. \\}

%==============
\collaboration{MAGIC Collaboration}
\author{V.~A.~Acciari}
\affiliation{Inst. de Astrof\'isica de Canarias, E-38200 La Laguna, and Universidad de La Laguna, Dpto. Astrof\'isica, E-38206 La Laguna, Tenerife, Spain}
\author{S.~Ansoldi} \affil{Universit\`a di Udine, and INFN Trieste,
I-33100 Udine, Italy} \affil{Japanese MAGIC Consortium: ICRR, The University of Tokyo, 277-8582 Chiba, Japan; Department of Physics, Kyoto University, 606-8502 Kyoto, Japan; Tokai University, 259-1292 Kanagawa, Japan; RIKEN, 351-0198 Saitama, Japan}
\author{L.~A.~Antonelli} \affil{National Institute for Astrophysics (INAF), I-00136 Rome, Italy} 
\author{A.~Arbet Engels} \affil{ETH Zurich, CH-8093 Zurich, Switzerland} 
\author{D.~Baack} \affil{Technische Universit\"at Dortmund, D-44221 Dortmund, Germany} 
\author{A.~Babi\'c} \affil{Croatian MAGIC Consortium: University of Rijeka, 51000 Rijeka, University of Split - FESB, 21000 Split,  University of Zagreb - FER, 10000 Zagreb, University of Osijek, 31000 Osijek and Rudjer Boskovic Institute, 10000 Zagreb, Croatia} 
\author{B.~Banerjee} \affil{Saha Institute of Nuclear Physics, HBNI, 1/AF Bidhannagar, Salt Lake, Sector-1, Kolkata 700064, India} 
\author{U.~Barres de Almeida}
\affil{Max-Planck-Institut f\"ur Physik, D-80805 M\"unchen, Germany}
\affil{now at Centro Brasileiro de Pesquisas F\'isicas (CBPF), 22290-180 URCA, Rio de Janeiro (RJ), Brasil}
\author{J.~A.~Barrio} \affil{Unidad de Part\'iculas y Cosmolog\'ia (UPARCOS), Universidad Complutense, E-28040 Madrid, Spain} 
\author{J.~Becerra Gonz\'alez} \affil{Inst. de Astrof\'isica de Canarias, E-38200 La Laguna, and Universidad de La Laguna, Dpto. Astrof\'isica, E-38206 La Laguna, Tenerife, Spain}
% ...


\begin{abstract}

Starburst galaxies  and  star-forming active galactic nuclei  (AGN) are among the  candidate sources  thought to contribute appreciably to the  extragalactic gamma-ray and neutrino backgrounds.
\textsc{NGC 1068} is the  brightest of the star-forming galaxies found to emit gamma rays from  0.1  to 50 GeV. 
 Precise measurements of the  high-energy spectrum  are crucial to study the particle accelerators and  probe the dominant emission mechanisms. 
We have carried out 125 hours of observations  of  \textsc{NGC 1068}  with the MAGIC telescopes in order to search for gamma-ray emission in the very high energy band.  We did not detect significant gamma-ray emission, and set upper limits at 95\% confidence level to the gamma-ray flux above  200 GeV  $f<$5.1$ \times$10$ ^{-13} $ cm$ ^{-2} $ s$ ^{-1} $. This limit improves previous constraints by about an order of magnitude and allows us to put tight constraints on the theoretical models for the gamma-ray emission.
By combining the MAGIC observations with the  {\it Fermi}-LAT spectrum we  limit the parameter space (spectral slope, maximum energy) of the cosmic ray protons predicted  by  hadronuclear models  for  the gamma-ray emission, while we find that a model postulating leptonic emission from a semi-relativistic jet is fully consistent with the limits.
We provide predictions for  IceCube detection of the neutrino signal foreseen in the hadronic scenario.  We predict a maximal IceCube neutrino  event rate of   0.07 yr$ ^{-1} $.

\end{abstract}




\section{Introduction} 

The cumulative gamma-ray and neutrino emission from star-forming galaxies, including starbursts and  star-forming active galactic nuclei (AGN), have been proposed to contribute to the extragalactic gamma-ray and neutrino backgrounds   \citep[e.g.][]{Tamborra14,Wang16_gamma, Lamastra17,Liu18}.
However, their  exact contributions to  the diffuse fluxes measured by the Large Area Telescope (LAT) on board the {\it Fermi Gamma-ray Space Telescope (Fermi)}  \citep{Ackermann15} and IceCube \citep{Aartsen15}  still have to be established.
Due to observational uncertainties in the measured spectra, the exact emission mechanisms and their parameters remain unknown. 

The gamma-ray emission in star-forming galaxies is expected  to be predominantly produced  from Cosmic Ray (CR) interactions with gas.  In these astrophysical environments CRs accelerated by supernova (SN) remnants interact with the interstellar medium (ISM) and produce neutral and charged  pions which in turn decay into high energy gamma rays and neutrinos   \citep[e.g.][]{Persic08,Rephaeli10,Yoast14, Eichmann16}.
Starburst galaxies  exhibit higher star formation rate (SFR$ \simeq $10-100 M$_{\odot}$ yr$ ^{-1} $) compared  to quiescently star-forming galaxies such as our Galaxy  (SFR$ \simeq $1-5 M$_{\odot}$ yr$ ^{-1} $, \citealt{Smith78,Murray10}).  Given the expected CR energy input from SN explosions and the dense gas present in starburst nuclei, starburst galaxies are expected to be more powerful gamma-ray emitters than normal  star-forming galaxies.
The starburst mode of star formation is likely  triggered by galaxy interactions (major and minor mergers), as suggested by  observational evidence and theoretical arguments 
\citep[e.g.][]{Sanders96, Hernquist89, Somerville01,Lamastra13a}.
 Consequently, galaxy interactions  also enhance the accretion of gas  into the central supermassive black hole and the ensuing AGN activity.
 The latter is also associated with galaxies undergoing secular evolution. Studies on the star-forming properties of AGN host galaxies indicate that 
the level of star formation in AGN hosts can be either elevated, as in starbursts, or normal, as in quiescently star-forming galaxies, or suppressed, as in passive spheroids (see \citealt{Lamastra13b, Gatti15, Rodighiero15} and references therein).
 In  active galaxies,  non-thermal radiation in the  gamma-ray band   may also be produced by the interaction of particles (protons and electrons) accelerated  in AGN-driven outflows (wind and  jet) with the ISM and interstellar radiation fields  \citep[e.g.][]{Lenain10,Tamborra14,Lamastra16, Lamastra19}.
 Indeed,  weak misaligned radio jet, and  wide-angle AGN-driven outflows  have  been observed in star-forming AGN detected in the GeV band by {\it Fermi}-LAT
      \citep{Gallimore96,Garcia14,Zschaechner16,Elmouttie98}. 
 The AGN contribution to gamma-ray emission  is supported by  the comparison between the galaxy non-thermal luminosity  and the  CR  luminosity  provided by star formation.  In starburst galaxies a  fraction equal to  $ \sim $0.3-0.6   of CR energy input is estimated to be converted into radiation in the gamma-ray band, while calorimetric fractions close to one,  and even larger,  have been observed in  star-forming AGN \citep{Ackermann12,WangFields16}.

The gamma-ray spectra of nearby starbursts and AGN  have been measured  in the high energy (HE, 0.1-100 GeV) band by {\it Fermi}-LAT  \citep{Ackermann12, Acero15,Fermi_3FHL,Lamastra16,Wojaczynski17,Hayashida13,Tang14,Peng16}.  The  starburst galaxies NGC 253 and M 82 have  also been detected in the very high energy (VHE, 0.1-100 TeV) band with Imaging Atmospheric Cherenkov Telescopes (IACTs) \citep{Acciari09,Acero09}. These two measurements are compatible and indicate  that the  gamma-ray spectra  can  be described by a single power-law with spectral index $ p \sim $2.2  up to TeV energies.

In this paper we present observations in the VHE band of the Seyfert galaxy \textsc{NGC 1068}   with the Major Atmospheric Gamma-ray Imaging Cherencov (MAGIC) telescopes. 
 \textsc{NGC 1068}, located at a distance of  D=14.4 Mpc, was detected  in the gamma-ray band  by   {\it Fermi}-LAT  \citep{Ackermann12,Fermi_3FHL}.  The latest spectral analysis based on 8 years of survey data in the 50 MeV - 1 TeV range  yields  a power-law index of $\sim$2.4 and  flux integrated  between 1 GeV and 100 GeV of 5.8$ \times$10$ ^{-13} $ cm$ ^{-2} $ s$ ^{-1} $ \citep{4FGL_cat}. 
Observations at  lower frequencies, have revealed the presence of both starburst and AGN activities. Interferometric observations in the millimetre  band have identified a $\sim$2 kpc starburst ring that surrounds a central molecular disk  of 350 pc$\times$200 pc size  in which a sizeable fraction of the  gas content is  involved in a massive AGN-driven  wind \citep{Garcia14,Krips11}. AGN-driven  jets  on scales from  hundreds  pc to kpc have been observed in the radio band   \citep{Gallimore96}.

Given the very different particle acceleration sites in an active galaxy, the exact origin of the measured high-energy gamma-ray emission in  \textsc{NGC 1068} is still undetermined.

The gamma-ray spectra predicted by  the starburst, AGN jet, and  AGN wind  models that have been proposed in literature differ significantly in the VHE band where IACTs are more sensitive than {\it Fermi}-LAT.  The leptonic  AGN jet model is characterized by a sharp cut-off at energies $ \sim $100 GeV , while the hadronic starburst and AGN wind models extend to the VHE  band, but with different spectral slopes.

In order to constrain the competing models, we conducted deep observations (125 hours) of NGC 1068 with the MAGIC telescopes. A detection or an upper limit on the VHE cut-off of the gamma-ray spectrum may provide valuable informations on the physical properties of the CR accelerator(s), and on  the  emission mechanism(s).
In particular, understanding the leptonic or hadronic nature of the gamma-ray  emission, and  the estimate of the maximum energy of accelerated particles  have important implications for the neutrino signal expected from this source.

The paper is organized as follow. In Section \ref{MAGIC_analysis}  we  present the MAGIC observations and data analysis. In Section \ref{Results} we  show the gamma-ray spectrum of   \textsc{NGC 1068} obtained by combining {\it Fermi}-LAT and MAGIC observations, and we derive constraints on the theoretical models for the gamma-ray emission.  In Section \ref{Discussion} we discuss the implications of the results, including that related to the neutrino signal. Conclusions follow in Section \ref{Conclusions}.

\section{MAGIC observations and  analysis}\label{MAGIC_analysis}

MAGIC is a stereoscopic system of two 17-m diameter IACTs situated at the Roque de los Muchachos, on the Canary island of La Palma (28.75$ ^{\circ}$N, 17.86$ ^{\circ}$W) at a height of 2200 m above sea level.
\textsc{NGC 1068}  observations were carried out from  January 2016 to January 2019 at zenith angles between 28$^{\circ}$ and 50$^{\circ}$,  in wobble mode \citep{Fomin94}, with a standard wobble offset of 0.4$^{\circ}$. 

Observations were taken under different night sky background (NSB) conditions.
Under dark night conditions, and for zenith angles $ < $30$^{\circ}$, MAGIC reaches a trigger energy threshold of $ \sim $50 GeV, and a sensitivity above 220 GeV of 0.67$\pm$0.04\% of the Crab nebula flux in 50 hours of observations \citep{MAGIC_sens}.
 The main effect of  moonlight  is an increase in the analysis energy threshold\footnote{The analysis energy threshold is obtained by fitting  the true energy distribution, which is  obtained by re-weighting the events with a power-law spectrum of index $p=$-2,  with a Gaussian function around its energy peak. We adopt $p=$-2 in the spectral analysis of the VHE data  because  we expect  a component of the gamma-ray emission harder than that measured in the HE band with {\it Fermi}-LAT.} and in the systematic uncertainties on the flux normalization. 
In order to limit the degradation of the energy threshold and of the sensitivity below 10\% we selected data samples that were recorded with nominal setting and with  NSB$<$8$\times$NSB$_{dark}$ ($ \sim$ 90\%  of the whole data sample).  After selecting the data with an aerosol transmission measured to be above 85\% that of a clear night, the final sample consists to a total of $\sim$ 125 hours of effective observation time of good data.

The data have been analysed using the standard MAGIC Analysis and Reconstruction Software (MARS), according to the prescriptions given in \cite{moon_paper}.
The recorded shower images were calibrated, cleaned and parametrized according to \cite{Hillas1985}  for each telescope individually. 
The analysis was performed using appropriate  Monte Carlo-simulated gamma-ray and background data  to reproduce the  observational conditions in each NSB data sample.  
The data reduction (stereo reconstruction, gamma/hadron separation, and estimation of energy and  arrival direction of the primary particle)  was performed for each sample separately. 
 The energy threshold, which is obtained by taking into account the actual zenith angle distribution of the selected data,  ranges from a minimum of about 140 GeV for the lowest NSB data sample to a maximum of about 300 GeV for the highest NSB data sample.
%The combined energy threshold was then calculated as the highest of the energy threshold derived for each sample.
Flux upper limits are calculated  following \cite{Rolke05}, with a confidence level of 95\%, and  considering a systematic error on flux estimation of 30\% \citep{MAGIC_sens}.

\section{Results}\label{Results}

\begin{figure}
\begin{center}
\includegraphics[width=9 cm]{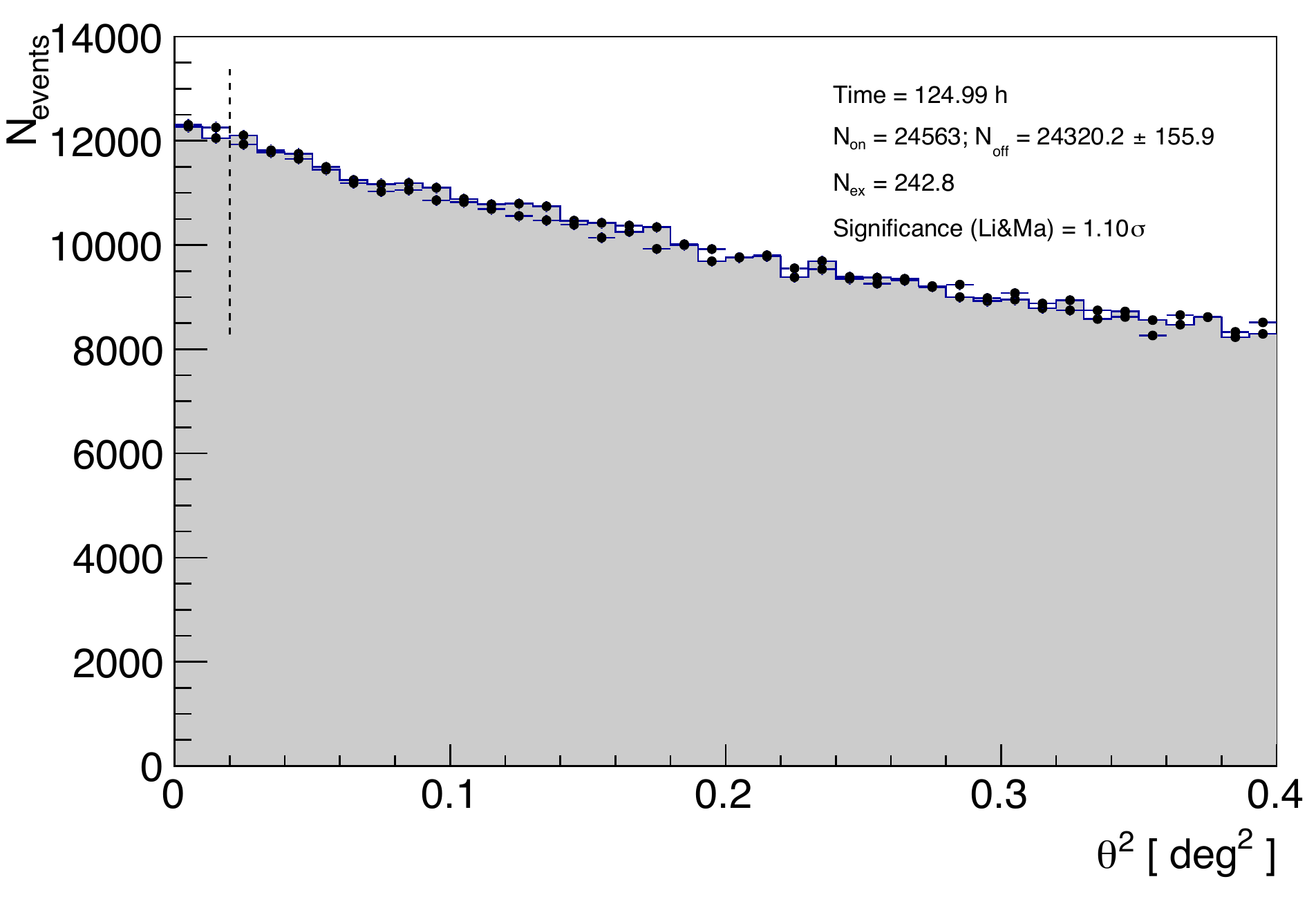}
\caption{Distribution of the square of the difference between the nominal position of the source and the reconstructed direction in camera coordinates for both the gamma-like events (blue crosses) and background events (grey histogram). The vertical dashed line marks the limit of the signal region.}
\label{NGC1068_odie}
\end{center} 
\end{figure}

Figure \ref{NGC1068_odie} shows the distribution of the square of the difference between the nominal position of the source and the reconstructed direction in camera coordinates for both the gamma-like events and background events.
For the total  dataset ($ \sim $125 hours), we find an excess of 243  gamma-like events over 24320$\pm$156  background events which yields a significance of  1.1 $ \sigma $ \citep{LiMa83}.  

 By excluding dataset with energy threshold larger than 200 GeV ($\sim$23 hours),  we derive an integral flux upper limit at 95\% confidence level above 200 GeV of  $f<$5.1$\times$10$^{-13}$ cm$^{-2}$ s$^{-1}$. This limit is  lower by about an order of magnitude than the  previous estimate  by \cite{Aharonian05}.  The latter is obtained from  4.3 hours of  observations  with the High Energy Stereoscopic System (H.E.S.S.) and with a slightly larger energy threshold of 210 GeV.  

\begin{figure*}
\begin{center}
\includegraphics[width=10cm]{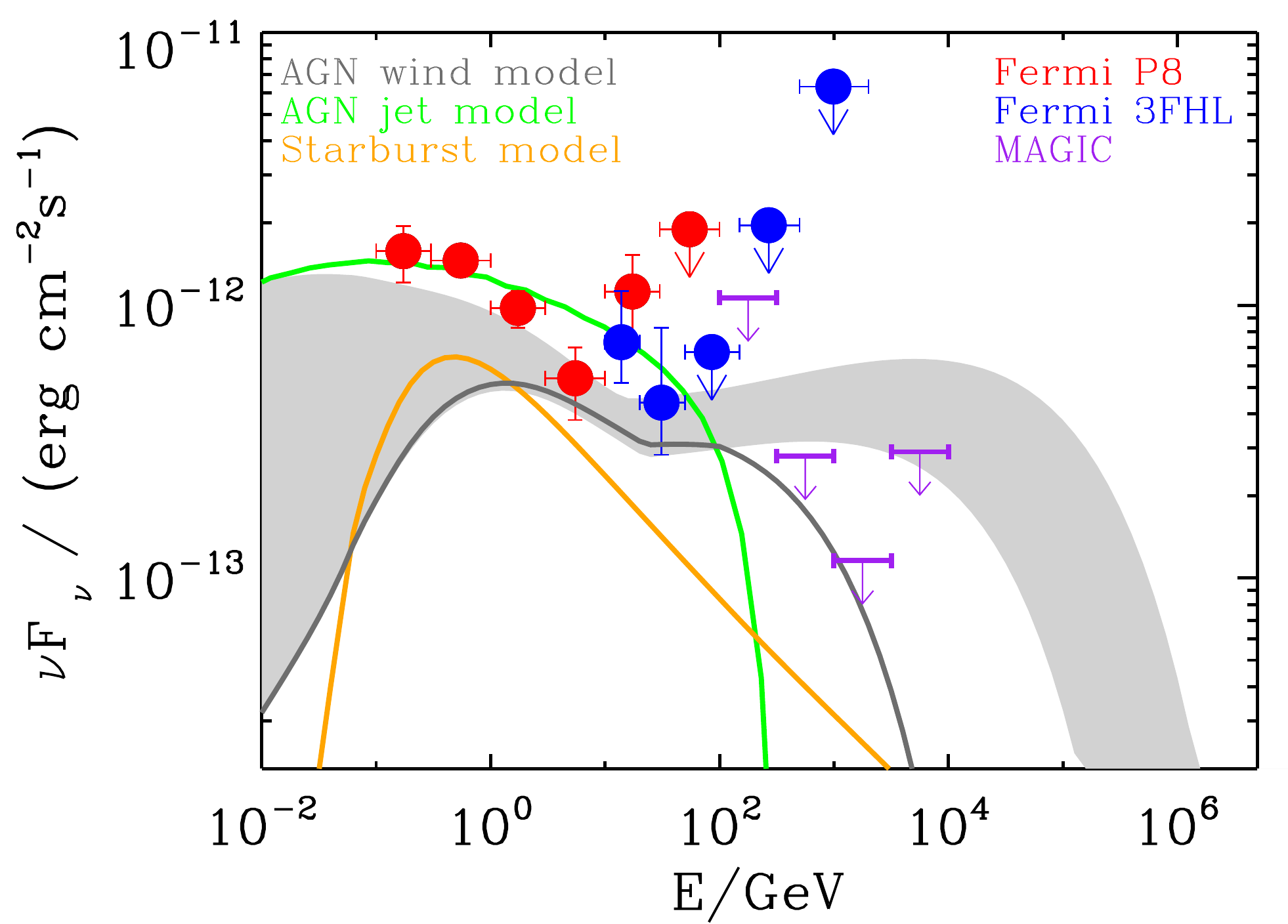}
\caption{Gamma-ray spectrum of \textsc{NGC 1068} in the HE and VHE band. The {\it Fermi}-LAT  data points are  from  \citealt{Lamastra16} (P8), and from  \citealt{Fermi_3FHL} (3FHL). The purple arrows indicate upper limits at 95\% confidence level derived from the analysis of MAGIC data ($ \sim $125 hours) presented in this paper. The green and orange lines show the gamma-ray spectra prediced by  the AGN jet (\citealt{Lenain10}) and starburst ($p$=2.5, $E_{\rm cut}$=10$ ^{8} $ GeV, and  $\xi$=0.04, \citealt{Eichmann16}) models, respectively.  The shaded grey band  indicates the upper ($p$=2, $E_{\rm cut}$=6$\times$10$ ^{6} $ GeV, and $\xi$=0.25) and lower ($p$=2, $E_{\rm cut}$=3$\times$10$ ^{5} $ GeV, and  $\xi$=0.2) bounds of the gamma-ray emission predicted by the AGN wind model  as proposed by \citealt{Lamastra16}.  For the sake of clarity, the predictions of the revised AGN wind model (\citealt{Lamastra19}) are not shown, since they do not differ from that by \citealt{Lamastra16}  at energies smaller than  10 TeV. For comparison,  the spectrum predicted by the  AGN wind model that is obtained by assuming  one of the combinations of CR proton spectral parameters compatible with the MAGIC upper limits ($p$=2, $E_{\rm cut}$=8$\times$10$ ^{3} $ GeV, and  $\xi$=0.2, see Figure \ref{gamma_Emax}) is shown with the dark grey line.  
 }
\label{NGC1068_spec}
 \label{bestfit}
\end{center} 
\end{figure*}

The differential flux upper limits in the VHE band obtained from the full data sample, as well as the  energy spectrum  measured with {\it Fermi}-LAT at lower energies \citep{Acero15,Lamastra16,Fermi_3FHL}, are shown in Figure \ref{NGC1068_spec} (see also Table 1). The gamma-ray emission was detected up to $ \sim $30 GeV  by  {\it Fermi}-LAT, while at higher energies only upper limits on the energy flux are determined. At energies $\sim$10 GeV an indication of a bump can be seen  in  the   {\it Fermi}-LAT spectrum. This apparent spectral feature could be ascribed  to the energy binning. As discussed in \cite{Lamastra16},  assuming a single bin in the (10-100) GeV  energy range a constant spectrum,  such as the one reported  in the 3FGL {\it Fermi} catalogue \citep{Acero15},  is obtained above 1 GeV.

In Figure \ref{NGC1068_spec},  we also show the spectra predicted by the starburst, AGN jet, and AGN wind models which have  been proposed to explain the gamma-ray emission  \citep{Eichmann16, Lenain10, Lamastra16}.  

The comparison between the predicted and observed gamma-ray spectra  indicates that 
the AGN jet models is in agreement with the observed gamma-ray flux and upper limits. In this model a maximum Lorentz factor of jet leptons of $\gamma_{max}$=10$^{6}$ is assumed in order to produce the sharp cut-off at $ \sim$100 GeV. On contrast, the AGN wind model predicts a hard spectrum  extending to the VHE band  that is strongly constrained by the MAGIC observations presented in this paper. Finally, the starburst model by \cite{Eichmann16}, where the gamma-ray emission is produced within the inner $ \sim $180  pc of the galaxy, is compatible with the VHE limits but cannot describe the {\it Fermi}-LAT spectrum; the gamma-ray flux at 1 GeV is higher than the model by about a factor of two.

The constrained part of the spectrum predicted by the AGN wind model  is the hadronic  component that  originates from the decay of neutral pions produced  in inelastic collisions between protons accelerated by  the  AGN-driven  outflow  observed in the molecular disk on $ \sim$100 pc scale and ambient protons.
The leptonic gamma-ray emission predicted by the AGN  wind model, as well as that predicted by the AGN jet model,  do not extend  at TeV energies owing to the effect of transition of IC  cooling from the Thomson regime to the Klein-Nishina regime. Thus, the limits on the VHE emission  can  be used to effectively constrain only  the hadronic gamma-ray emission  of the AGN wind and starburst models.

To derive constraints on the  CR proton population of star formation and of AGN wind origin,   we compare the gamma-ray spectra predicted by the starburst and AGN wind models with the spectrum measured in the HE band and  with the upper limits derived in the VHE band.
In both the starburst and AGN wind models protons are assumed to be  accelerated by diffusive shocks with an energy distribution  $N(E)=AE^{-p}\exp(-E/E_{cut})$, where the normalisation constant $A$ is determined by the total energy supplied to relativistic protons at the shock,   $p\simeq$2 is the spectral index, and $E_{cut}$ is the maximum energy of accelerated protons. The latter  has a physical maximum limit determined by the Hillas criterion: $E_{max}=10^{18}Z(R/kpc)(B/\mu G)$eV, where $Z$ is the atomic charge number, $R$ is the physical extent of the acceleration region, and $B$ is the magnetic field \citep{Hillas1985}, while the minimum energy of accelerated protons is  the proton rest mass.    

With regard to the  the energy input from star formation, since  {\it Fermi}-LAT does not spatially resolve the gamma-ray emitting region, we consider  the total  star formation of the  galaxy. The kinetic input from  star formation  is calculated as  
$L_{kin}^{\rm SF}=\nu_{\rm SN}E_{\rm SN}$, where $\nu_{\rm SN}$ is the supernovae rate, and $E_{\rm SN}\simeq 10^{51}$ erg is the typical kinetic energy from a supernova explosion. We estimated $\nu_{\rm SN}$=0.43 yr$ ^{-1} $ from the total infrared luminosity of the galaxy  L$_{\rm IR} \simeq 10^{45}$ erg s$^{-1}$  (between 8 and  1000 $\mu$m, \citealt{Ackermann12}), and  assuming  a Kroupa initial mass function \citep{Kroupa01,Kennicutt12}, yielding  $L_{kin}^{\rm SF}$=1.4$\times10^{43}$ erg s$^{-1}$.  We find that the kinetic luminosity  provided by the  star formation throughout  the galaxy can  produce the gamma-ray emission measured in the  {\it Fermi}-LAT band.

As regard the AGN wind model, we derived the kinetic luminosity provided by the AGN  from  the kinetic luminosity of the  molecular outflow which is observed by millimetre interferometers on $ \sim $100 pc scale, yielding   $L^{\rm AGN}_{\rm kin}=(0.5-1.5)\times10^{42}$ erg s$^{-1}$ \citep{Krips11,Garcia14,Lamastra16}. This molecular outflow is likely produced by the  interaction of the molecular  gas with either the AGN jet, and/or the energy released during accretion of matter onto the supermassive black hole,  rather than star formation\footnote{The  star formation rate SFR$ \simeq $1 M$_{\odot}$/yr  for the circumnuclear region up to a radius  $R\simeq$140 pc  \citep{Esquej14} is unable to power the molecular outflow.}.

The fraction of the  CR energy input provided by star formation and AGN activity  that is emitted  in gamma rays depends on the  proton acceleration efficiency, $ \eta$,  and on the efficiency of converting proton kinetic energy into gamma rays,  $F_{cal}$. The comparison between the gamma-ray  emission in star-forming galaxies   and the  kinetic energy supplied by  SN explosions leads to   $\eta\simeq$(0.1-0.3) and    $F_{cal} \simeq $0.3-0.6   \citep{Keshet03,Tatischeff08,Lacki10,Ackermann12,WangFields16}.

\begin{figure*}
\begin{center}
\includegraphics[width=8 cm]{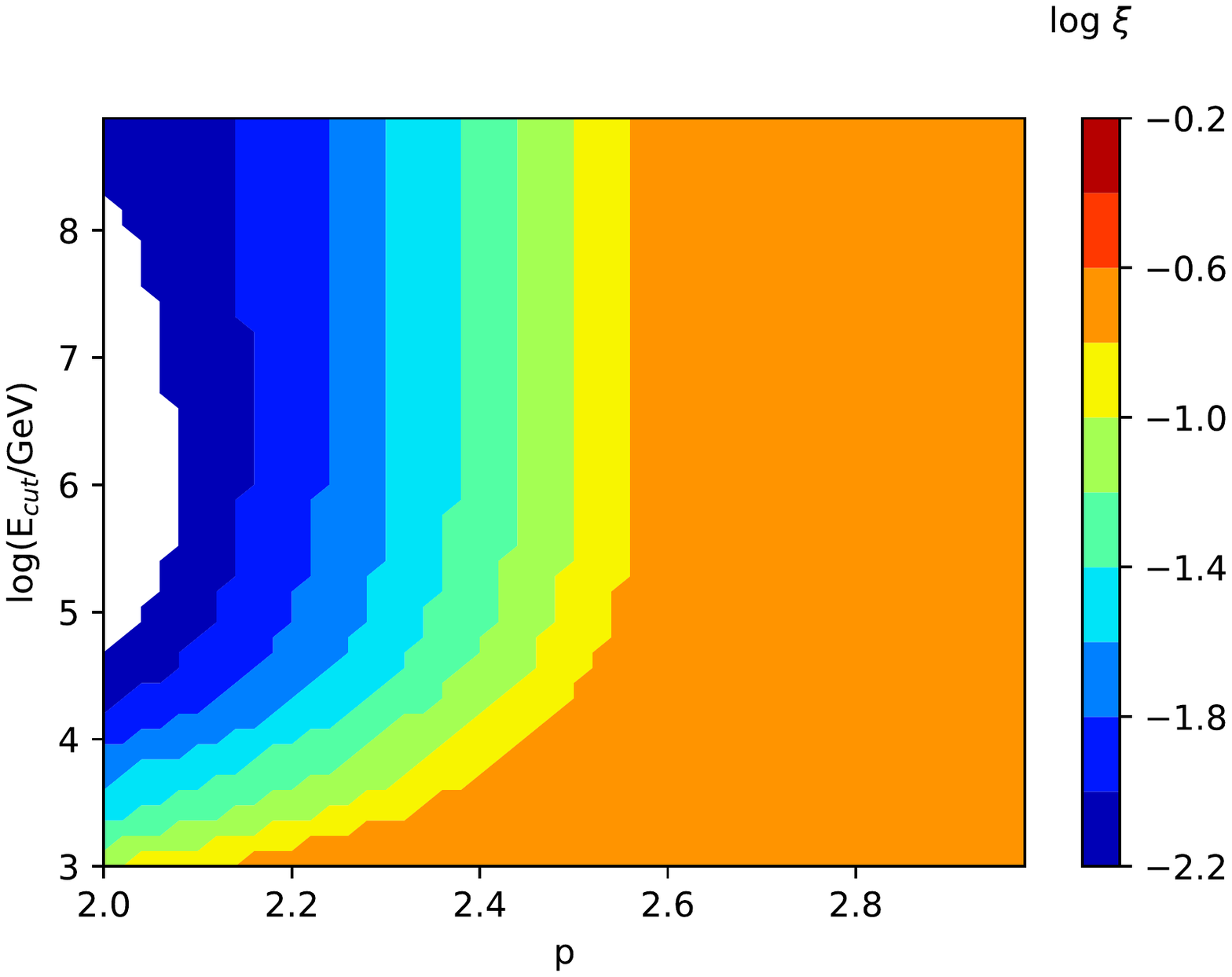}
\includegraphics[width=8 cm]{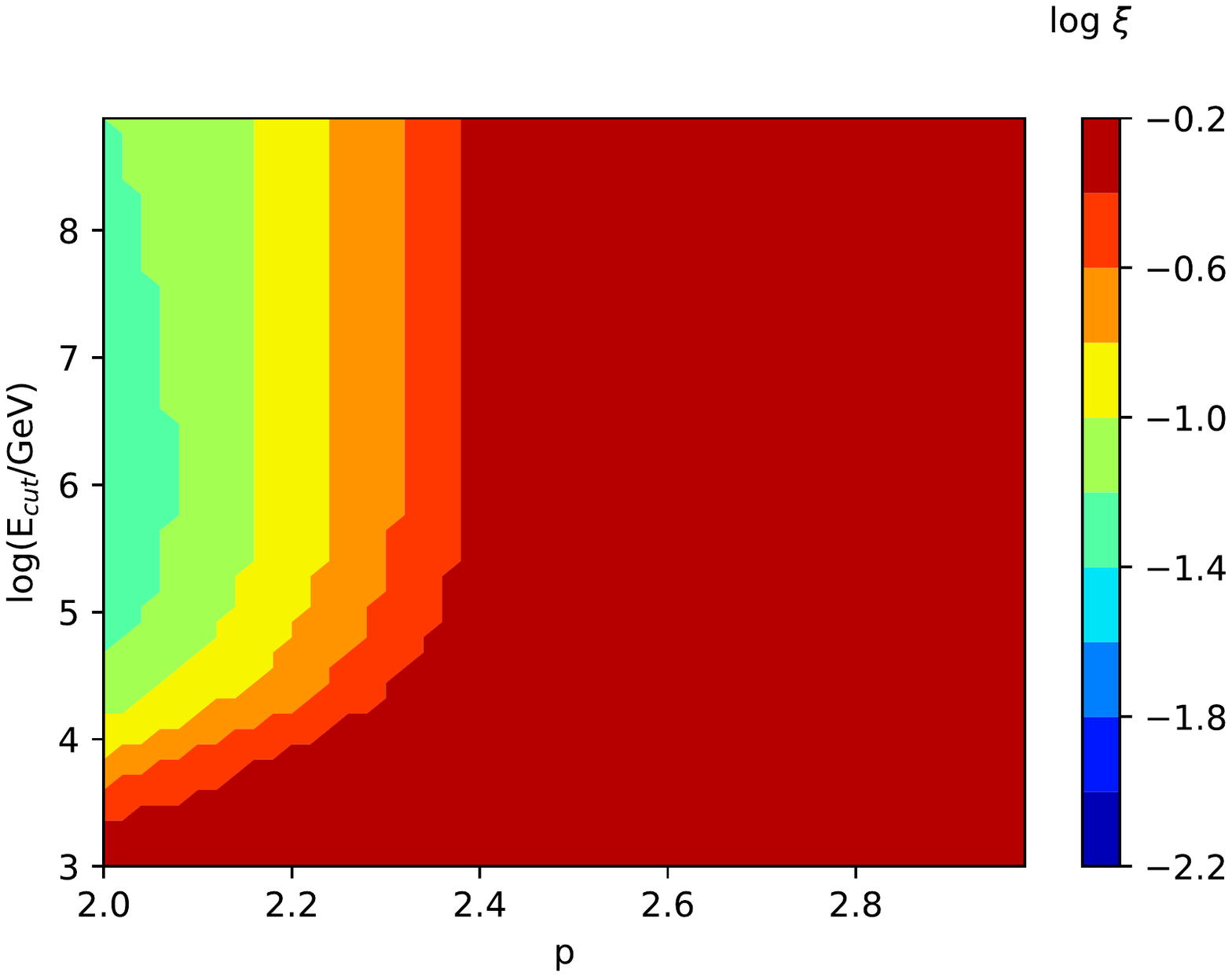}
\caption{The allowed values of the starburst (left) and AGN wind (right) hadronic  spectral index $p$ and cut-off energy $E_{cut}$ derived from HE and VHE observations of  \textsc{NGC 1068} are plotted as a function of the product between the calorimetric fraction and the acceleration efficiency $\xi$.  The logarithmic values of $\xi$ corresponding to the coloured contours are displayed on the vertical bar. } 
\label{gamma_Emax}
\end{center} 
\end{figure*}

We compute the  gamma-ray spectrum produced by  neutral pion decays  following proton-proton interactions  as in \cite{Lamastra16} (see also \citealt{Kelner06}), and varying   the CR proton parameters:  $p$, $E_{\rm cut}$, and  $\xi=\eta \times F_{\rm cal}$.  The comparison between the predicted spectra and the upper limits in the VHE band allowed us to derive reliable constraints on these parameters.  The results  are shown in Figure \ref{gamma_Emax} where for each  value of  $\xi$, the allowed value of $E_{\rm cut}$ is plotted as a function of $p$ in the starburst and AGN models, separately.
In each model we find that,  for a given value of $\xi$,  the cut-off energy increases with the spectral index  up to a value at which $E_{\rm cut}$ becomes  independent on  the value of $p$.
This is determined by the gradually lower effect of the high-energy cut-off on the shape and normalization of the gamma-ray spectrum, depending on the fraction $\xi$ and slope $p$.

The constraints obtained for the  CR proton parameters  are mainly determined by the MAGIC upper limits in the  0.3-1 TeV  and 1-3 TeV energy bins.  Gamma rays with energy above the threshold for electron-positron pair production may be absorbed due to  interactions with  the extragalactic background light    (EBL, e.g. \citealt{Dominguez2011:EBL, Franceschini17,Acciari19}), and  radiation fields within the galaxy.
Both processes contribute in turn to the  uncertainty in the computation of $p$ and $E_{\rm cut}$. 
As discussed in \cite{Lamastra19},   absorption due to the EBL
% \footnote{To calculate the EBL absorption we assume a  distance of \textsc{NGC 1068} of D=14.4 Mpc} 
and to the infrared emission from the starburst ring surrounding the acceleration regions   affects the gamma-ray spectrum only above  $E \gtrsim$10 TeV, thus the constraints shown in Figure \ref{gamma_Emax} can be considered robust.

\begin{deluxetable*}{cc}[b!]
\tablecaption{Spectral energy distribution in VHE band.\label{tab1}}
\tablecolumns{2}
%\tablenum{2}
%\tablewidth{0pt}
\tablehead{
\colhead{log E} &
\colhead{$\nu F_{\nu}$ }  \\
\colhead{(GeV) } &
\colhead{ ( erg cm$^{-2}$ s$^{-1}$)}  \\
}
\startdata
2.25 & $<$1.1$\times$10$^{-12}$ \\
2.75  &$<$ 2.8 $\times$10$^{-13}$  \\
3.25  & $<$1.1 $\times$10$^{-13}$ \\
3.75 &$<$ 2.9 $\times$10$^{-13}$ \\
\enddata
%\tablenotetext{a}{At exposure start.}
%\tablecomments{}
\end{deluxetable*}

\section{Discussion}\label{Discussion}

The derived properties of the gamma-ray spectrum  of   \textsc{NGC 1068} are now compared  with the gamma-ray properties of the two starburst galaxies detected in the VHE band.  
NGC 253 and M 82 were detected by  H.E.S.S.  and VERITAS in $ \simeq $180 hours and $ \simeq $ 140   hours of observations, respectively.
After the detection at TeV energies, these starburst galaxies were also detected by  {\it Fermi}-LAT. 
Although the statistic is limited, there is no indication of a spectral break or cut-off features apparent in the spectra. The smooth alignment of the GeV and TeV spectrum suggests that a single energy loss mechanism dominates in the gamma-ray band.  
The loss mechanism is probably related to CR proton  interactions with ISM. 
 In fact, models assuming that a population of protons is giving rise to the measured gamma-ray spectra through hadronic collisions provide  a good fit to the data \citep{Abramowski12}. Moreover, the lack of variability  at any gamma-ray energy supports the idea that the emission is related to star formation processes rather than nuclear activity.
The gamma-ray properties of \textsc{NGC 1068}, including the shape of the gamma-ray spectrum in the HE band and the lack of variability of the {\it Fermi}-LAT  and MAGIC lightcurves, suggest that the radiation processes are similar to those in the other starburst galaxies.  
However, the extrapolation of  the gamma-ray emission of \textsc{NGC 1068} in the  VHE  band assuming a single power law with spectral index  $p\simeq$2.2, as resulting from  the combined fit of the latest  {\it Fermi}-LAT and H.E.S.S. spectrum of NGC 253 \citep{NGC253_HESS_18}, results in  an over-prediction of the  emission   at TeV energies.  As  shown in  Figure \ref{gamma_Emax},  in the starburst model  a gamma-ray spectrum that extends to energies  $E_{\rm cut}\gtrsim$10$ ^{4} $ GeV, with  $p=$2.2 can only be obtained  for low calorimetric and  acceleration efficiencies ($\xi\lesssim $0.02,) that produce  a gamma-ray flux at 1 GeV lower by about a factor of ten than that measured by  {\it Fermi}-LAT.
In  case the emission in the {\it Fermi}-LAT band is ascribed to star formation,  loss mechanisms that make the starburst spectrum soft and/or the proton  maximum energy low, must operate in the starburst region.

The precise measurements of the gamma-ray spectral properties of starburst galaxies, including those with AGN, are crucial to determine their contribution to the extragalactic gamma-ray and neutrino backgrounds.  The estimates of the  source  population contribution  to the observed backgrounds   rely on the extrapolation of the characteristic source spectrum to the region between the HE band and the IceCube energy scale. Analyses that have utilized GeV-TeV gamma-ray spectral information to constrain the contribution of starburst galaxies  to the diffuse fluxes measured by {\it Fermi}-LAT and IceCube found that a characteristic  CR proton spectral index of $p\simeq$2.2   and cut-off energy $E_{\rm cut}$=10$ ^{7} $ GeV  are consistent with the bounds from the residual non-blazar component of the extragalactic gamma-ray background. This yields a contribution to the diffuse neutrino background of 30\% at 100 TeV, and 60\% at 1 PeV \citep{Bechtol17}. A harder ($p\simeq$2.1) spectrum saturates the IceCube signal, while a  softer ($p\simeq$2.3) spectrum underestimates the diffuse neutrino energy flux by about an order of magnitude, remaining compatible with the gamma-ray bounds \citep{Linden17,Bechtol17,Palladino18}.

\begin{figure}
\begin{center}
\includegraphics[width=8 cm]{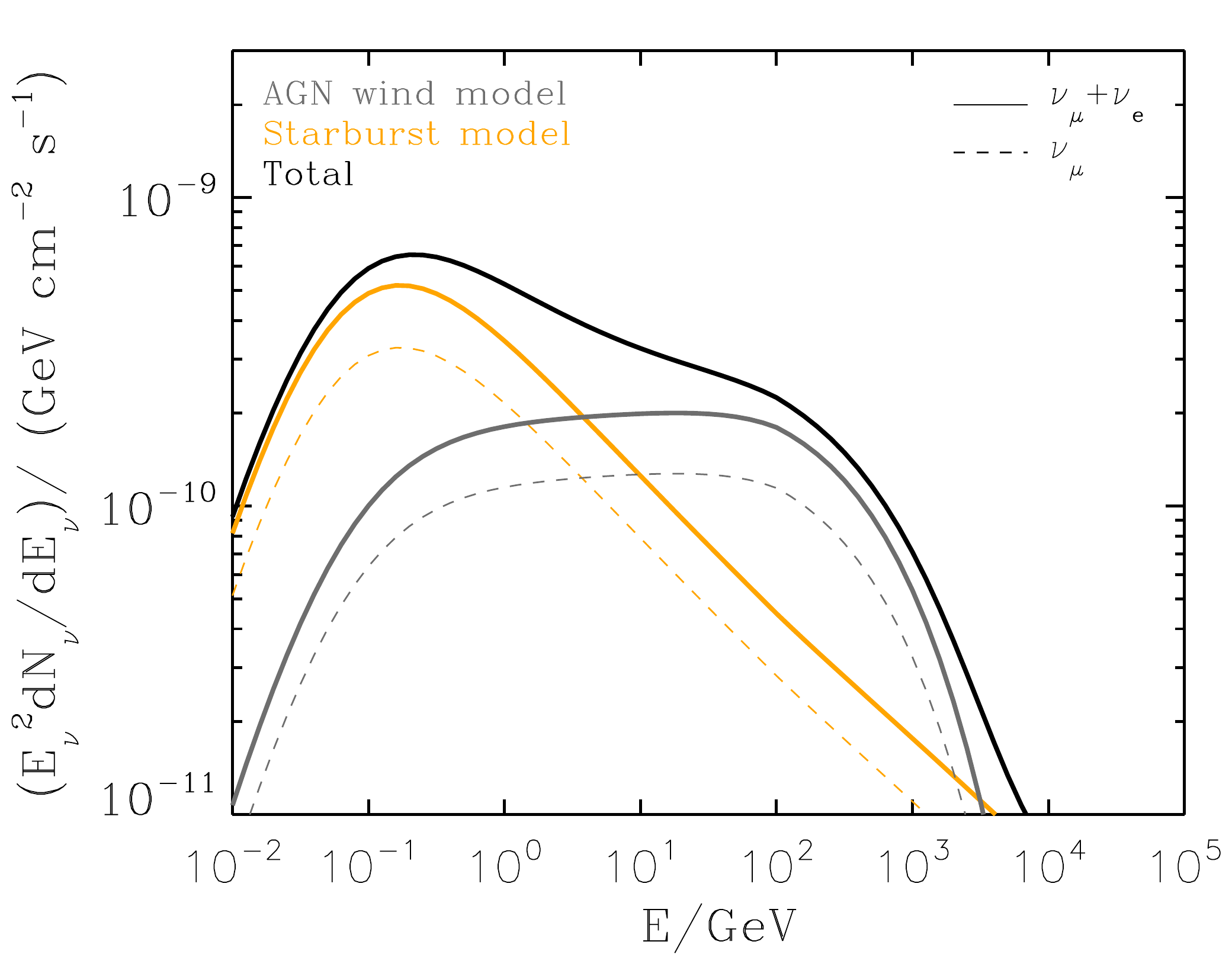}
\caption{ Neutrino spectra of \textsc{NGC 1068} expected in the starburst (orange lines) and  AGN wind (grey lines) models. Dashed lines indicate the muon neutrino fluxes, while solid lines indicate the total neutrino fluxes.  The solid black line represents the cumulative neutrino spectrum. }
\label{NGC1068_neutrino}
\end{center} 
\end{figure}

We apply this multi-messenger approach to  \textsc{NGC 1068}  and we derive the expected neutrino flux  based on the observed gamma-ray flux.
%In this context  it is useful to determine the expected neutrino flux of \textsc{NGC 1068}  based on the observed gamma-ray flux.
In Figure \ref{NGC1068_neutrino}  we show the neutrino spectra  predicted by the starburst model and by the  AGN wind model with CR parameters compatible with VHE upper limits  that are shown in Figure \ref{bestfit}.  
To calculate the neutrino spectra  we use the parametrisations for  the high-energy spectra of secondary particles produced in proton-proton collisions derived by \cite{Kelner06}.
We calculate the spectra of muon and electron (anti)neutrinos  from muon decays, and the spectra of muon (anti)neutrinos  produced through the direct decays of charged pions.

In order to assess the capability of current neutrino detectors in  testing  the hadronuclear models, following \cite{Lamastra16},
we combine the total neutrino spectra with the effective area of IceCube  \citep{Aartsen14}. We obtain a neutrino event rate  with energy $E_{\nu}>$0.1 TeV  of  0.002  $ yr^{-1} $ and 0.001 $ yr^{-1}$  for  the starburst  and  the AGN wind models, respectively. 
 Besides, we compute the maximum IceCube neutrino event rates, compatible with the MAGIC upper limits, scanning the parameter space of the starburst and AGN wind models, described in Figure  \ref{gamma_Emax}. We obtain a maximum event rate of 0.07  yr$ ^{-1} $.

The level  of neutrino signal predicted  by hadronuclear models makes the detection  of \textsc{NGC 1068}   a challenge  for the current neutrino detectors.  A neutrino flux  larger than those derived in this paper can be achieved if the source of neutrinos resides in extremely dense environment which prevents the escape of GeV-TeV gamma rays.  This would  argue for gamma-ray production  in the AGN core, where the intense optical and  near infrared emission produced by the active nucleus and the surrounding dusty torus \citep{Honig08} could act as  the target photon field for both photohadronic gamma-ray and neutrino emissions and  for pair production   \citep[e.g.][]{Murase16,inoue19}.

\section{Conclusions}\label{Conclusions}

The results from the MAGIC observations of  \textsc{NGC 1068} imply that the gamma-ray spectrum could be  either entirely produced by leptonic processes, as in the AGN jet model,  or, if  a hadronuclear component is present, as envisaged in the AGN wind or in the starburst models, the accelerated proton population should have soft  spectra ($p\gtrsim$2.2) and/or  low maximum energy ($E_{\rm cut}\simeq$10$ ^{4}$ GeV).
 
 At present, it is not possible to resolve spatially  the emission from the different components (jet, starburst, molecular disk)  in the gamma-ray band with  {\it Fermi}-LAT, thus no strong conclusions can be drawn on their relative contributions to the observed emission. 
 This obstacle could be overcome in principle  with observations in the radio band  that can potentially benefit also from  spatial information. However, the presence of the  radio jet in the inner 100 pc  hampers the identification of any emission not originating from the jet or the compact nucleus.  
Firm conclusions cannot  be drawn on the different contributions on the basis of the non variability of the gamma-ray flux. In fact,  the gamma-ray emission in the AGN jet model  may be  produced  from  a few tenth of parsecs up to hundred parsec from the nucleus \citep{Lenain10}, and no significant  variability is expected for the more distant emitting zone, as in the starburst and AGN wind models.

Improving our understanding of the gamma-ray spectral properties of  star forming galaxies and AGN is crucial to test source population models of the extragalactic gamma-ray and neutrino backgrounds. 
Indeed, although coincident observations of neutrinos and gamma rays from the blazar  TXS 0506+056  represent a compelling evidence of the first extragalactic neutrino source \citep{Neutrino_paper, MAGIC_neutrino}, independent analyses indicate that blazars can account only for $ \lesssim$30\% of the diffuse neutrino flux measured by IceCube \citep{Padovani16,Murase16b,IceCube17}. 

The astrophysical high-energy neutrino  flux observed with  IceCube is consistent with an isotropic distribution of neutrino arrival directions, suggesting an extragalactic origin.   
Star-forming galaxies  such as  \textsc{NGC 1068}, could be the main contributors to the observed neutrino emission. 
The  increase of the sensitivity up to a factor $ \sim $10, as envisaged in the the next generation of neutrino detectors (such as Km3Net and IceCube-Gen2), will allow the detection of neutrinos from the starburst and AGN-wind scenarios described here. The detection of a neutrino signal from these sources would be  a compelling evidence for the presence of a hadronic component in the gamma-ray spectrum.
At the same time, the improved sensitivity of the next generation of ground based gamma-ray observatories, like the Cherenkov Telescope Array, will allow us to disentangle the different emission mechanisms. In particular, simulations of 50 hours of observations of \textsc{NGC 1068} with CTA have shown that leptonic and hadronic  models could be distinguished \citep{Lamastra19}.

\acknowledgments

We would like to thank the  Referee for useful comments and  
the Instituto de Astrof\'{\i}sica de Canarias for the excellent working conditions at the Observatorio del Roque de los Muchachos in La Palma. The financial support of the German BMBF and MPG, the Italian INFN and INAF, the Swiss National Fund SNF, the ERDF under the Spanish MINECO (FPA2015-69818-P, FPA2012-36668, FPA2015-68378-P, FPA2015-69210-C6-2-R, FPA2015-69210-C6-4-R, FPA2015-69210-C6-6-R, AYA2015-71042-P, AYA2016-76012-C3-1-P, ESP2015-71662-C2-2-P, FPA2017‐90566‐REDC), the Indian Department of Atomic Energy, the Japanese JSPS and MEXT, the Bulgarian Ministry of Education and Science, National RI Roadmap Project DO1-153/28.08.2018 and the Academy of Finland grant nr. 320045 is gratefully acknowledged. This work was also supported by the Spanish Centro de Excelencia ``Severo Ochoa'' SEV-2016-0588 and SEV-2015-0548, and Unidad de Excelencia ``Mar\'{\i}a de Maeztu'' MDM-2014-0369, by the Croatian Science Foundation (HrZZ) Project IP-2016-06-9782 and the University of Rijeka Project 13.12.1.3.02, by the DFG Collaborative Research Centers SFB823/C4 and SFB876/C3, the Polish National Research Centre grant UMO-2016/22/M/ST9/00382 and by the Brazilian MCTIC, CNPq and FAPERJ.



\end{document}